\documentclass[aps,prd,twocolumn,groupedaddress,showpacs]{revtex4}

\usepackage{graphicx}
\usepackage{color}
\usepackage[T1]{fontenc}

\newcommand{\beIV}{\mbox{$^7$Be$^{3+}$}}

\begin{document}


\title{$^7$Be charge exchange between \beIV\ ion and exotic long-lived negatively charged massive particle in big bang nucleosynthesis}


\author{Motohiko Kusakabe$^{1,2}$}
\email{motohiko@kau.ac.kr} 
\author{K. S. Kim$^{1}$}
\email{kyungsik@kau.ac.kr} 
\author{Myung-Ki Cheoun$^{2}$}
\email{cheoun@ssu.ac.kr} 
\author{Toshitaka Kajino$^{3,4}$}
\email{kajino@nao.ac.jp}
\author{Yasushi Kino$^{5}$}
\email{y.k@m.tohoku.ac.jp}
\affiliation{
$^1$School of Liberal Arts and Science, Korea Aerospace University, Goyang 412-791, Korea}
\affiliation{
$^2$Department of Physics, Soongsil University, Seoul 156-743, Korea}
\affiliation{
$^3$National Astronomical Observatory of Japan, 2-21-1 
Osawa, Mitaka, Tokyo 181-8588, Japan}
\affiliation{
$^4$Department of Astronomy, Graduate School of Science, 
University of Tokyo, 7-3-1 Hongo, Bunkyo-ku, Tokyo 113-0033, Japan}
\affiliation{
$^5$Department of Chemistry, Tohoku University, Sendai 980-8578, Japan
}


\date{\today}

\begin{abstract}
The existence of an exotic long-lived negatively charged massive particle, i.e., $X^-$, during big bang nucleosynthesis can affect primordial light element abundances.  Especially, the final abundance of $^7$Li, mainly originating from the electron capture of $^7$Be, has been suggested to reduce by the $^7$Be destruction via the radiative $X^-$ capture of $^7$Be followed by the radiative proton capture of the bound state of $^7$Be and $X^-$ ($^7$Be$_X$).  We suggest a new route of $^7$Be$_X$ formation, that is the $^7$Be charge exchange at the reaction of \beIV\ ion and $X^-$.  The formation rate depends on the number fraction of \beIV\ ion, the charge exchange cross section of \beIV\, and the probability that produced excited states $^7$Be$_X^\ast$ are converted to the ground state.  We estimate respective quantities affecting the $^7$Be$_X$ formation rate, and find that this reaction pathway can be more important than ordinary radiative recombination of $^7$Be and $X^-$.  The effect of the charge exchange reaction is then shown in a latest nuclear reaction network calculation.  Quantum physical model calculations for related reactions are needed to precisely estimate the efficiency of this pathway in future.
\end{abstract}


\maketitle


\section{Introduction}

New physics operating during the big bang nucleosynthesis (BBN) can be probed by observed light element abundances.  Possible indications of new physics come from discrepancies between primordial  abundances of $^6$Li and $^7$Li predicted in standard BBN (SBBN) model and those inferred from observations of metal-poor stars (MPSs).  These MPSs exhibit a plateau-like abundance ratio, $^7$Li/H $=(1-2) \times 10^{-10}$ at low  metallicities of [Fe/H]$>-3$~\cite{Spite:1982dd,Ryan:1999vr,Melendez:2004ni,Asplund:2005yt,bon2007,shi2007,Aoki:2009ce,Hernandez:2009gn,Sbordone:2010zi,Monaco:2010mm,Monaco:2011sd,Mucciarelli:2011ts,Aoki:2012wb,Aoki2012b}, and much lower at extremely low metallicities of [Fe/H]$<-3$~\cite{Frebel:2005ig,Aoki:2005gn} \footnote{[A/B]$=\log(n_{\rm A}/n_{\rm B})-\log(n_{\rm A}/n_{\rm B})_\odot$, where $n_i$ is the number density of $i$ and the subscript $\odot$ indicates the solar value, for elements A and B.}.  The plateau abundance is a factor of 2--4 lower than the SBBN prediction for the baryon-to-photon ratio determined from the observation of the cosmic microwave background radiation with {\it Wilkinson Microwave Anisotropy Probe} (WMAP) (e.g.,
$^7$Li/H=$(5.24^{+0.71}_{-0.67})\times 10^{-10}$~\cite{Cyburt:2008kw}).  This discrepancy
indicates a need of some mechanism to decrease the $^7$Li abundance.  Although astrophysical processes such as the combination of atomic and turbulent diffusion in stellar atmospheres~\cite{Richard:2004pj,Korn:2007cx} may be a cause of the observed abundances, this is not yet established~\cite{Lind:2009ta}.  

In recent spectroscopic observations of MPSs, the lithium isotopic ratio of $^6$Li/$^7$Li is also measured.  A possible plateau abundance of $^6$Li/H$\sim 6\times10^{-12}$ has been suggested~\cite{Asplund:2005yt}, which is about 1000 times higher than the SBBN prediction.  Since an effect of convective motions in stellar atmospheres could cause asymmetries in atomic line profiles and consequently leads to an erroneous estimation of $^6$Li abundance~\cite{Cayrel:2007te}, the effect should be estimated.  Even including this effect, high $^6$Li abundances have been evidently detected in at most several MPSs \cite{asp2008,gar2009,ste2010,ste2012}.   Such a high abundance level at the low-metallicity requires processes other than standard Galactic cosmic-ray (CR) nucleosynthesis models \cite{pra2006}.

To investigate the $^{6,7}$Li problems, in this work, we focus on a model to resolve the lithium problems assuming the presence of negatively charged massive particles $X^-$~\cite{cahn:1981,Dimopoulos:1989hk,rujula90} with a mass much larger than the nucleon mass, i.e., $m_X \gg 1$ GeV, during the BBN epoch~\cite{Pospelov:2006sc,Kohri:2006cn,Cyburt:2006uv,Kaplinghat:2006qr,Hamaguchi:2007mp,Bird:2007ge,Kusakabe:2007fu,Kusakabe:2007fv,Jedamzik_nega,Kamimura:2008fx,Pospelov:2007js,Kawasaki:2007xb,Jittoh_nega,Pospelov:2008ta,Khlopov:2007ic,Kawasaki:2008qe,Bailly:2008yy,Kamimura2010,Kusakabe:2010cb,Pospelov:2010hj,Kohri:2012gc,Cyburt:2012kp,Dapo2012}.
One of candidates for the $X^-$ particle is stau, the supersymmetric partner of tau \cite{Jittoh_nega,Cyburt:2006uv}.  The $X^-$ particles become electromagnetically  bound to positively charged nuclides with binding energies of $\sim
O(0.1-1)$~MeV.  The bound state of a nuclide $A$ and an $X^-$ particle is denoted by $A_X$.  Because of these low binding energies the bound states cannot form until late in the BBN epoch, when nuclear
reactions are no longer efficient. Hence, the effect
of the $X^-$ particles is rather small \cite{Kusakabe:2007fv}.  Depending upon their abundance and lifetime, however, the
$X^-$ particles can affect lithium abundances through the following processes:  1) $^6$Li production via the recombination of $^4$He with $X^-$ followed by the catalyzed $\alpha$ transfer reaction $^4$He$_X$($d$,$X^-$)$^6$Li~\cite{Pospelov:2006sc,Hamaguchi:2007mp,Dapo2012}, 2) $^7$Be destruction via the recombination of $^7$Be with $X^-$ followed by the radiative proton capture reaction $^7$Be$_X$($p$,$\gamma$)$^8$B$_X$.  In the latter process the destruction of $^7$Be occurs through both the first atomic excited state of $^8$B$_X$~\cite{Bird:2007ge} with excitation energy of 0.819 MeV in the limit of infinite $X^-$ mass \cite{Kamimura:2008fx} or the atomic ground state (GS) of $^8$B$^\ast$($1^+$,0.770~MeV)$_X$ composed of the $1^+$ nuclear excited state of $^8$B and an $X^-$~\cite{Kusakabe:2007fu}~\footnote{In SBBN with the baryon-to-photon ratio inferred from WMAP,  $^7$Li is produced mostly as $^7$Be during the BBN epoch, which transforms into $^7$Li by electron capture in late universe.}.  

In all previous investigations, all processes are assumed to start from radiative recombination of fully ionized nuclides $A$ and $X^-$ as suggested in Ref. \cite{Dimopoulos:1989hk,rujula90}.  However, we found a possibility that nonradiative charge exchange reactions between hydrogen-like ions and $X^-$ can contribute to change nuclear abundances in the BBN model including an $X^-$ particle.  We focus on the $^7$Be nuclide since the effect on its abundance is expected to be important.  We introduce a simple model for estimating the significance of the new process in Sec.~\ref{sec2}, and show the result of $^7$Be$_X$ formation rate and time evolution of nuclear abundances in example cases in Sec.~\ref{sec3}.  Finally we summarize this work in Sec.~\ref{sec4}.  We adopt natural units, $\hbar=c=k_{\rm B}=1$, where $\hbar$ is the reduced Planck constant, $c$ is the speed of light, and $k_{\rm B}$ is the Boltzmann constant.

\section{model}\label{sec2}

\subsection{Effective rate for nonradiative recombination of $^7$Be and $X^-$}\label{sec21}
We suggest a process of $^7$Be$_X$ production via the \beIV\ ion formation.  This process can be important since $^7$Be$_X$ is destroyed through the reaction $^7$Be$_X$($p$, $\gamma$)$^8$B$_X$ and the final abundance of $^7$Li changes.   The destruction of $^7$Be$_X$ occurs at the temperature $T_9\equiv T/(10^9~{\rm K})\sim 0.4$ or $T\sim 34$ keV \cite{Bird:2007ge,Kusakabe:2007fv}, so that we focus on this temperature region.  In this epoch, there are two nonradiative reactions producing $^7$Be$_X^\ast$.  They are the substitution reaction \beIV\ +$X^- \rightarrow ^7$Be$_X^\ast+e^-$, and the three body collisional recombination $^7$Be$+X^-+e^\pm \rightarrow ^7$Be$_X^\ast$+$e^\pm$, where the ground and excited states of fully ionized $^7$Be$^{4+}$ and $X^-$ are denoted by $^7$Be$_X$ and $^7$Be$_X^\ast$ as usual, while the partially ionized hydrogen-like ion is denoted by $^7$Be$^{3+}$ throughout this work.  The latter reaction is not considered in this work since the number abundance of $e^\pm$ is not so large in this late epoch of $e^\pm$ pair annihilation and consequently the reaction rate is not large in this temperature region.

The effective rate for $^7$Be to recombine with $X^-$ particle through the \beIV\ ion formation is given by the product of the following three quantities: 1) the number ratio of \beIV\ to $^7$Be$^{4+}$ (=$^7$Be) which is the dominant chemical species containing the $^7$Be nuclide, 2) the formation rate of excited states $^7$Be$_X^\ast$ via the reaction $^7$Be$^{3+}$+$X^-\rightarrow ^7$Be$_X^\ast+e^-$, and 3) the probability of converting excited states $^7$Be$_X^\ast$ to the GS $^7$Be$_X$ which is estimated by the ratio of the transition rate producing the GS $^7$Be$_X$ and the total reaction rate of the excited state $^7$Be$_X^\ast$.  The effective recombination rate is then described as
\begin{eqnarray}
\Gamma_{\rm rec}&=&\frac{n_{{\rm Be}^{3+}}}{n_{{\rm Be}^{4+}}} \left[\Gamma_{{\rm Be}^{3+}\rightarrow {\rm Be}_X^\ast} \frac{\Gamma_{{\rm Be_X^\ast},~{\rm tr}}}{\Gamma_{{\rm Be_X^\ast},~{\rm de}}+\Gamma_{{\rm Be_X^\ast},~{\rm tr}}}\right],
\label{eq18}
\end{eqnarray}
where
$n_i$ denotes the number density of species $i=$Be$^{3+}$ and Be$^{4+}$, 
$\Gamma_{{\rm Be}^{3+}\rightarrow {\rm Be}_X^\ast}$ represents the rate for \beIV\ to form excited states $^7$Be$_X^\ast$. 
$\Gamma_{{\rm Be}_X^\ast,~{\rm tr}}$ and $\Gamma_{{\rm Be}_X^\ast,~{\rm de}}$ are rates for the transition to the GS and the destruction of excited states $^7$Be$_X^\ast$, respectively.  The parameters in the square bracket have uncertainties related to the cross sections for formation and destruction of $^7$Be$_X^\ast$, and the fraction of the produced $^7$Be$_X^\ast$ which decays to the GS $^7$Be$_X$ through multiple electric dipole transitions.  A precise estimation of these uncertainties is, however, beyond the scope of this study.  

\subsection{Hydrogen-like ion}\label{sec22}
The energy level for the main quantum number $n$ is
\begin{equation}
E_n=-\frac{Z_i^2 \alpha^2 \mu(i, j)}{2n^2},
\label{eq7}
\end{equation}
where
$Z_i$ is the proton number of particle $i$ ($Z_{^7{\rm Be}}=4$),
$\alpha$ is the fine structure constant, and 
$\mu$($i$, $j$) is the reduced mass for the $i$+$j$ two body system.  The charge number of particle $j=e^-$ or $X^-$ was assumed to be $-1$.
The binding energy for $n$ is related to the energy level as $E_{\rm B}(n)=-E_n$.
The expectation value of radius $r$ between the $i$ and $j$ particles composing the hydrogen-like ion is
\begin{equation}
\langle r \rangle =\frac{n^2}{Z_i \alpha \mu(i, j)} \left[1+\frac{1}{2}\left(1- \frac{\ell(\ell+1)}{n^2}\right)\right],
\label{eq8}
\end{equation}
where
$\ell$ is the azimuthal quantum number.  There is a relation between the average radius and the binding energy from Eqs. (\ref{eq7}) and (\ref{eq8}):
\begin{equation}
\langle r \rangle \sim \frac{n^2}{Z_i\alpha \mu(i, j)} = -\frac{Z_i\alpha}{2E_n}
\label{eq9}
\end{equation}
The average radius is thus inversely proportional to the binding energy independent of the reduced mass of the system.  

The binding energy of the GS \beIV\ is given by Eq. (\ref{eq7}) to be 217.7 eV.  In the limit of $m_X\gg 1$ GeV, the reduced mass for the system of $^7$Be and $X^-$ is given by $\mu(^7{\rm Be}, X^-)\rightarrow m_{^7{\rm Be}}=6.53$ GeV.  The main quantum number of $^7$Be$_X^\ast$ state whose binding energy is nearly equal to that of the GS \beIV\ is $n_{\rm eq}\sim$113.  We note that the reaction $^7$Be$^{3+}$+$X^-\rightarrow ^7$Be$_X^\ast+e^-$ is similar to the reaction H+$\bar{p}\rightarrow p\bar{p}+e^-$, where $p\bar{p}$ is the protonium consisting of a proton and an antiproton \cite{Cohen1997,Sakimoto2001,Cohen2004} since the following three characters are shared by the two reactions.  Firstly, components related to the reaction are two heavy particles and a light electron.  The heavy particles are $^7$Be and $X^-$ in the former reaction, and $p$ and $\bar{p}$ in the latter.  Secondly, a charged heavy particle ($^7$Be and $p$, respectively) is transfered.  Thirdly, a bound state of heavy two particles ($^7$Be$_X$ and $p\bar{p}$, respectively) is produced.  In the case of the protonium formation, $p\bar{p}$ states with the main quantum numbers of $n_{\rm eq}\sim$30 have binding energies nearly equal to that of the hydrogen GS (13.6 eV).  The reaction of the muonic hydrogen formation, H$+\mu^- \rightarrow p\mu+e^-$ \cite{Cohen2004,Sakimoto2010}, is also a similar reaction although the muon mass, 106 MeV, is lighter than the nucleon mass by one order of magnitude.

\subsection{The ground state of \beIV\ is an isolated system}\label{sec23}
The average radius of the \beIV\ GS is given by Eq. (\ref{eq8}) as
\begin{equation}
\langle r_{\rm 1S} \rangle = \frac{3}{2Z_{^7{\rm Be}}\alpha \mu(^7{\rm Be},~e^-)} =1.98\times 10^{-9}~{\rm cm}.
\label{eq1}
\end{equation}
The reduced mass $\mu$($^7$Be, $e^-$) is approximately given by the electron mass $m_e$.

The average distance between $e^\pm$'s at the temperature $T=34$ keV is given by
\begin{equation}
l_{\rm ave}=n_{e^\pm}^{-1/3}=4.5\times 10^{-8}~{\rm cm},
\label{eq2}
\end{equation}
where
$n_{e^\pm}$ is the number density of $e^\pm$.  We assumed that the number density is given by the formula for the non-degenerate non-relativistic fermion \cite{Kolb:1990vq}, i.e.,
\begin{eqnarray}
n_{e^\pm}&=&g_e\left(\frac{m_e T}{2\pi}\right)^{3/2} \exp\left(-\frac{m_e}{T}\right),\nonumber\\
&=&1.1\times 10^{22}~{\rm cm}^{-3}~~~({\rm for}~T=34~{\rm keV}),
\label{eq3}
\end{eqnarray}
where
$g_e=2$ is the statistical degrees of freedom both for electron and positron.
Since the average distance is larger than the average radius of the \beIV\ GS, the GS can be regarded as an isolated two-body system.

\subsection{Balance of \beIV\ ionization and $^7$Be$^{4+}$ recombination: $^7$Be$^{4+}+e^-\rightleftharpoons$ \beIV\ $+\gamma$}\label{sec24}
We adopt the rate for the GS \beIV\ formation via the radiative recombination of $^7$Be and $e^-$ from Ref. \cite{Verner:1995br}:
\begin{equation}
\alpha_{\rm r}(T)=a \left[\sqrt[]{\mathstrut \frac{T}{T_0}} \left(1+\sqrt[]{\mathstrut \frac{T}{T_0}}\right)^{1-b} \left(1+\sqrt[]{\mathstrut \frac{T}{T_1}}\right)^{1+b}\right]^{-1},
\label{eq4}
\end{equation}
where parameter values are
$a=4.290\times 10^{-10}$ cm$^3$ s$^{-1}$, $b=0.7557$, $T_0=30.00$ K, and $T_1=1.093\times 10^7$ K.  Although this is the total rate for transitions to all final states including not only the GS but also excited states, the formation of the GS dominates for the temperature range of $T>10^8$ K \cite{Verner:1995br}.  The adoption of the rate is thus justified.  For the related temperature, the rate is $\alpha_{\rm r}(T_9=0.4)=5.13\times 10^{-16}$ cm$^3$ s$^{-1}$.  The rate for $^7$Be to recombine with $e^-$ is then given by $n_{e^-} \alpha_{\rm r}=5.77\times 10^6$ s$^{-1}$ at $T_9=0.4$.  This rate is much larger than the Hubble expansion rate, i.e., $H(T)=4.50\times 10^{-4}$ s$^{-1}$ $(T_9/0.4)^2$.  The recombination and its inverse reaction, i.e., the \beIV\ ionization, is thus very effective, and the number ratio of \beIV\ to $^7$Be$^{4+}$ is the equilibrium value.

The equilibrium number ratio is described by Saha equation:
\begin{eqnarray}
\frac{n_{{\rm Be}^{3+}}}{n_{{\rm Be}^{4+}} \cdot n_{e^-}} &=&\frac{g_{{\rm Be}^{3+}}}{g_{{\rm Be}^{4+}} \cdot g_{e}} \left(\frac{m_{{\rm Be}^{3+}}}{m_{{\rm Be}^{4+}} \cdot m_{e}}\right)^{3/2} \left(\frac{2\pi}{T}\right)^{3/2} \nonumber\\
&& \times \exp\left(\frac{m_{{\rm Be}^{4+}} +m_{e} -m_{{\rm Be}^{3+}}}{T}\right),
\label{eq5}
\end{eqnarray}
where
$g_i$ and $m_i$ are the spin degrees of freedom and the mass, respectively, of species $i=$Be$^{3+}$, Be$^{4+}$, and $e^-$.
The spin factors are $g_{{\rm Be}^{3+}}=2[2 I_{\rm s}({\rm Be})+1]$ (for the GS), $g_{{\rm Be}^{4+}}=2 I_{\rm s}({\rm Be})+1$, and $g_e=2$, respectively, with $I_{\rm s}({\rm Be})$ the nuclear spin of a given Be isotope.   Using the spin factors, the approximate equality of $m_{{\rm Be}^{3+}}\approx m_{{\rm Be}^{4+}}$, and Eq. (\ref{eq3}), the number ratio is derived as
\begin{eqnarray}
\frac{n_{{\rm Be}^{3+}}}{n_{{\rm Be}^{4+}}} &=&\left(\frac{2\pi}{m_e T}\right)^{3/2} \exp\left[\frac{I(^7{\rm Be}^{3+})}{T}\right]n_e\nonumber\\
&=&2\exp\left[\frac{I(^7{\rm Be}^{3+})-m_e}{T}\right]\sim 2 \mathrm{e}^{-m_e/T},
\label{eq6}
\end{eqnarray}
where
the ionization potential $I(i)$ was defined for species $i$, and $I(^7{\rm Be}^{3+})=m_{{\rm Be}^{4+}} +m_{e} -m_{{\rm Be}^{3+}}=217.7$ eV [Eq. (\ref{eq7})].
This ratio is $n_{{\rm Be}^{3+}}/n_{{\rm Be}^{4+}}=5.94\times 10^{-7}$ for $T_9=0.4$.

We note that the pair annihilation in the collision of $^7$Be$^{3+}$ and $e^+$, i.e., \beIV($e^+$, 2$\gamma$)$^7$Be is another process for the conversion from \beIV\ to $^7$Be.  However, effects of this reaction are negligible since its rate is much smaller than the rate for photoionization of \beIV.  The cross section of the annihilation for non-relativistic velocities $v$ is given by $\sigma_{\rm ann}\sim \pi \alpha^2/(m_e^2 v)$ \cite{Peskin1995}.  The annihilation rate is then roughly given by
\begin{eqnarray}
\Gamma_{\rm ann}&=&n_{e^+} \pi \alpha^2/m_e^2 \nonumber\\
&=& 8.2\times 10^7~{\rm s}^{-1}~~~(T=34~{\rm keV}),
\end{eqnarray}
[cf. Eq. (\ref{eq3})].  On the other hand, using the detailed balance relation \cite{Rybicki1979}, an lower limit on the photoionization rate of \beIV\ is derived as 
\begin{eqnarray}
\Gamma_{\rm ion}&>& \frac{g_{{\rm Be}^{3+}}}{g_{{\rm Be}^{4+}} \cdot g_{e}} \left[\frac{\mu(^7{\rm Be},~e^-)T}{2\pi}\right]^{3/2} \alpha_{\rm r}(T) \mathrm{e}^{-I(\beIV)/T} \nonumber\\
&\sim& 9.7 \times 10^{12}~{\rm s}^{-1}~\left[\frac{\mu(^7{\rm Be},~e^-)}{m_e}\right]^{3/2}~~~(T=34~{\rm keV}), \nonumber\\
\end{eqnarray}
where
the lower limit is derived by a replacement of the Planck distribution function of photon with the Boltzmann distribution function \cite{Pagel1997}.  The ionization rate is much larger than the annihilation rate, i.e., $\Gamma_{\rm ion} \gg \Gamma_{\rm ann}$.  The annihilation is thus insignificant, and neglected in this work.

The equilibrium abundance of $^7$Be$^{3+}$ ion decays even in this BBN epoch although the fraction of decaying $^7$Be$^{3+}$ is negligible.  The half life for electron capture of $^7$Be$^{3+}$ is $T_{1/2}=53.22$ day \cite{Be2004} =$4.60\times 10^6$ s.  It is much longer than the time scale of the universe corresponding to the $^7$Be$_X$ formation epoch, $1.78 \times 10^4$ s $(g_\ast/3.36)^{-1/2}(T_9/0.1)^{-2}$ with $g_\ast$ the effective number of relativistic degrees of freedom in terms of energy density \cite{Kolb:1990vq}.

\subsection{$^7$Be$^{3+}$+$X^-\rightarrow ^7$Be$_X^\ast+e^-$ reaction}\label{sec25}
The thermal reaction rate at a given temperature $T$ is given (Eq. (2.52) of Ref. \cite{Pagel1997}) by
\begin{equation}
\langle \sigma v \rangle =\left[\frac{8}{\pi \mu(i, j)}\right]^{1/2} \frac{1}{T^{3/2}} \int_0^\infty E \sigma(E) \exp\left(-\frac{E}{T}\right) dE,
\label{eq10}
\end{equation}
where
$E$ is the center of mass kinetic energy,
and $\sigma(E)$ is the reaction cross section as a function of $E$.
We assume that the cross section is similar to that of the protonium formation.  It is then roughly taken to be $\sigma(E)=\sigma(I(^7{\rm Be}^{3+}))[E/I(^7{\rm Be}^{3+})]^{-1/2} H(I(^7{\rm Be}^{3+})-E)$ with $\sigma(I(^7{\rm Be}^{3+}))$ the cross section at $E=I(^7{\rm Be}^{3+})$ and $H(x)$ the step function (Fig. 9 of Ref. \cite{Sakimoto2001}).  Since the ionization, i.e., $^7$Be$^{3+}+X^-\rightarrow ^7$Be$+X^-+e^-$, dominates for energies above the ionization threshold, the cross section for $E>I(^7{\rm Be}^{3+})$ is negligibly small.  Supposing the scaling of the cross section $\sigma\propto \langle r \rangle^2$, the cross section for the \beIV+$X^-$ reaction is assumed to be $\sigma(I(^7{\rm Be}^{3+}))=10/(Z_{^7{\rm Be}}\alpha m_e)^2=1.75\times 10^7$ b.

We note that the reaction of muonic hydrogen formation, $^1$H($\mu^-$, $e^-$)$p\mu$, is a same kind of reaction as the reaction $^7$Be$^{3+}$($X^-$, $e^-$)$^7$Be$_X^\ast$.  The muonic hydrogen formation is the first reaction occurring in experiments of muonic atoms \cite{Bertin1975,Cohen2004}.  Observables in the experiments such as energy flux densities of uncaptured muon and x-rays emitted from excited states of muonic atoms, however, do not provide a direct information on the muon capture rate.  Comparisons of theories and experiments on the cross section is, therefore, difficult for the moment \cite{Cohen2004}.  Then we can only refer to theoretical cross sections.  Calculations of the cross sections based on several different methods have been performed, and they agrees well at least qualitatively (Fig. 2 of Ref. \cite{Cohen2004}).  The cross sections $\sigma(E)$ scale as $\sim 10/(\alpha m_e)^2$ similarly to that for the protonium formation \footnote{This scaling as well as the amplitude of the cross section has been confirmed in a recent rigorous quantum mechanical calculation \cite{Sakimoto2010}.  Small amplitudes of undulation structures have been also found to exist in the $\sigma(E)$ curve which are caused by a quantum mechanical effect.}.

Under these assumptions, the reaction rate is given by
\begin{eqnarray}
\langle \sigma v \rangle &\approx&\frac{4\sqrt{\mathstrut 2}}{3\sqrt{\mathstrut \pi}} \frac{I(^7{\rm Be}^{3+})^2}{\mu(^7{\rm Be}^{3+}, X^-)^{1/2} T^{3/2}} \sigma(I(^7{\rm Be}^{3+})), \nonumber\\
&=& 5.23
\times 10^{-14}~{\rm cm}^3~{\rm s}^{-1} \left(\frac{T_9}{0.4}\right)^{-3/2} \left[\frac{\sigma(I(^7{\rm Be}^{3+}))}{1.75\times 10^7~{\rm b}}\right],\nonumber\\
\label{eq11}
\end{eqnarray}
where
we assumed that $\exp(-E/T)=1$ since the cross section is nonzero only for $E\lesssim T/160$.

The number density of baryon is given by
\begin{eqnarray}
n_b&\approx& \frac{\rho_b}{m_p}=\frac{\rho_c\Omega_b}{m_p}\left(1+z\right)^3\nonumber\\
&=&7.56
\times 10^{17}~{\rm cm}^{-3}\left(\frac{h}{0.700}\right)^2 \left(\frac{\Omega_b}{0.0463}\right)\left(\frac{T_9}{0.4}\right)^3,\nonumber\\
\label{eq12}
\end{eqnarray}
where
$\rho_b$ and $\rho_c$ are the baryon density and the present critical density, respectively.  
$\Omega_b=0.0463 \pm 0.0024$ is the baryon density parameter,
$z$ is the redshift of the universe which has a relation with temperature as $(1+z)=T/T_0$ with $T_0=2.7255$ K the present radiation temperature of the universe \cite{Fixsen:2009ug}, and
$h=H_0$/(100 km s$^{-1}$ Mpc$^{-1}$)$=0.700\pm 0.022$ is the reduced Hubble constant.  The cosmological parameters are taken from values determined from the WMAP \cite{Spergel:2003cb,Spergel:2006hy,Larson:2010gs,Hinshaw:2012fq} (Model $\Lambda$CDM, WMAP9 data only).

The rate for \beIV\ to form excited states of $^7$Be$_X^\ast$ through the charge exchange reaction is then given by
\begin{eqnarray}
\Gamma_{{\rm Be}^{3+}\rightarrow {\rm Be}_X^\ast}&=&n_X \langle \sigma v \rangle=n_b Y_X \langle \sigma v \rangle\nonumber\\
&=&3.96
Y_X \times 10^4~{\rm s}^{-1}~\left(\frac{h}{0.700}\right)^2 \left(\frac{\Omega_b}{0.0463}\right)\nonumber\\
&&\times \left(\frac{T_9}{0.4}\right)^{3/2} \left[\frac{\sigma(I(^7{\rm Be}^{3+}))}{1.75\times 10^7~{\rm b}}\right],
\label{eq13}
\end{eqnarray}
where
$n_X$ is the number density of the $X^-$ particle, and $Y_X=n_X/n_b$ is the number ratio of $X^-$ to baryon.

\subsection{Bound-bound transition of $^7$Be$_X^\ast$}\label{sec26}

\subsubsection{Dominant process}\label{sec261}

Excited states $^7$Be$_X^\ast$ produced via the $^7$Be charge exchange reaction experience bound-bound transitions including not only the spontaneous emission but also the stimulated emission and the photo-absorption in the early universe filled with the cosmic background radiation (CBR).  The spectrum of CBR in BBN epoch is assumed to be the Planck function, i.e.,
\begin{eqnarray}
B_\nu(T)=\frac{4\pi \nu^3}{\exp(2\pi \nu/T)-1},
\label{eq26}
\end{eqnarray}
where
$\nu=E_\gamma/(2\pi)$ is the frequency of the photon with $E_\gamma$ the photon energy.

Dominant process which occurs most frequently is determined as follows.  The transition from the upper ($u$) to the lower ($l$) energy states is considered.  Energy levels of the upper and lower states are defined by $E_u$ and $E_l$, respectively.  The energy and the frequency of emitted photon is then $E_{ul}=E_u-E_l$ and $\nu_{ul}=E_{ul}/(2\pi)$, respectively.  The spontaneous emission rate for a transition from $u$ to $l$ is defined with the Einstein $A$-coefficient, i.e., $A_{ul}$.  The photo-absorption rate and the stimulated emission rate are described with $B$-coefficients as $B_{lu}B_{\nu_{ul}}(T)$, and $B_{ul}B_{\nu_{ul}}(T)$, respectively.  In addition, the Einstein relations are given by
\begin{eqnarray}
g_l B_{lu} &=& g_u B_{ul}, \label{eq27} \\
A_{ul}&=&4\pi \nu^3 B_{ul}.
\label{eq28}
\end{eqnarray}
The ratio between rates for the stimulated emission and the spontaneous emission then satisfies
\begin{equation}
\frac{B_{ul}B_{\nu_{ul}}(T)}{A_{ul}}=\frac{1}{\exp(E_{ul}/T)-1},
\end{equation}
where
Eqs. (\ref{eq26}) and (\ref{eq28}) were used.  Clearly, transitions of $E_{ul} \lesssim T$ proceeds predominantly through the stimulated emission, while those of $E_{ul} \gtrsim T$ proceeds predominantly through the spontaneous emission.  The photo-absorption rate is equal to the stimulated emission rate except for the spin factors $g_l$ and $g_u$ [Eq. (\ref{eq27})].

\subsubsection{Spontaneous emission}\label{sec262}

The Einstein $A$-coefficient is given (Eq. (10.28b) of Ref. \cite{Rybicki1979}) by
\begin{eqnarray}
A_{ul}=\frac{4}{3} E_{ul}^3 \frac{1}{g_u} \sum \left|d_{ul}\right|^2,
\label{eq14}
\end{eqnarray}
where
$d_{ul}$ is the electric dipole matrix element for the initial state $u$ and the final state $l$, and
the sum is over all magnetic substates of the upper and lower states.  The difference in energy between levels with nearest main quantum numbers is $E_n-E_{n-1}=Z_i^2 \alpha^2 \mu(i, j) (n-1/2)/[n^2(n-1)^2]$ $\rightarrow Z_i^2 \alpha^2 \mu(i, j)/n^3$ (for $n\gg 1$).  Here the typical difference in energy between $u$ and $l$ is taken as $E_{ul}\sim |E_u|$ assuming that the main quantum numbers of $u$ ($n_u$), and $l$ ($n_l$) are close to each other.  

Here we introduce the absorption oscillator strength \cite{Rybicki1979} defined as
\begin{eqnarray}
f_{lu}=\frac{2m_A}{3 Z_A^2 \alpha} E_{ul} \frac{1}{g_l} \sum \left|d_{lu}\right|^2,
\label{eq24}
\end{eqnarray}
where
$m_A$ and $Z_A$ are the mass and proton number, respectively, of nuclide $A$ bound to $X^-$ (in the present case, $A=^7$Be).  Typical amplitude of this strength is known as $f_{lu} \lesssim {\mathcal O}(1)$ (see Refs. \cite{Karzas1961,Rybicki1979}).  We note that this definition is different from that for hydrogen-like electronic ions in terms of mass and charge number for the following reason.  In the harmonic oscillation model for the latter case, an electron with classical radius $r_{0,e}=e^2/m_e$ is bound to an nucleus.  In the case of $^7$Be$_X^\ast$, on the other hand, an $^7$Be nucleus with classical radius $r_{0,^7{\rm Be}}=(Z_{^7{\rm Be}} e)^2/m_{^7{\rm Be}}$ is bound to an $X^-$.  From Eqs. (\ref{eq14}) and (\ref{eq24}), we find
\begin{eqnarray}
A_{ul}=\frac{2 Z_A^2 \alpha}{m_A} \frac{g_l}{g_u} f_{lu} E_{ul}^2.
\label{eq25}
\end{eqnarray}

The number of final states is about $N_l\sim 2n$ including $\sim n$ levels for azimuthal quantum numbers of final states which are different from that of the initial state by 1 and $-1$, respectively.   The total spontaneous emission rate  of excited state $^7$Be$_X^\ast$ is then roughly estimated as
\begin{eqnarray}
\Gamma_{u,~{\rm sp}}&=&\sum_l A_{ul}\nonumber\\
&=& \frac{2N_l Z_A^2\alpha}{m_A} \overline{E_{ul}^2} \nonumber \\
&\simeq&5.83
\times 10^{11}~{\rm s}^{-1} \left(\frac{N_l}{226}\right) \left[\frac{\overline{E_{ul}^2}}{(218~{\rm eV})^2}\right]\nonumber\\
&&\times \left(\frac{Z_A}{4}\right)^2 \left(\frac{m_A}{6.53~{\rm GeV}}\right)^{-1},
\label{eq15}
\end{eqnarray}
where
the weighted average quantity $\overline{E_{ul}^2}\equiv \sum_l [(g_l/g_u)f_{lu} E_{ul}^2]/N_l$ was defined.

\subsubsection{Excitation and deexcitation by CBR}\label{sec263}

The Einstein coefficient $B_{ul}$ is given \cite{Rybicki1979} by
\begin{equation}
B_{ul}=\frac{4\pi^2 Z_A^2 \alpha}{E_{ul} m_A} \frac{g_l}{g_u}f_{lu}.
\label{eq22}
\end{equation}

The rate for the deexcitation via the stimulated emission is estimated to be
\begin{eqnarray}
\Gamma^{\gamma}_{u,~{\rm st}}&=& \sum_l B_{ul} B_{\nu_{ul}}(T) \nonumber \\
&=&\sum_l \frac{2 Z_A^2 \alpha}{m_A} \frac{g_l}{g_u} f_{lu} \frac{E_{ul}^2}{\exp(E_{ul}/T)-1} \nonumber \\
&\sim&\frac{2N_l Z_A^2 \alpha}{m_A} T \overline{E_{ul}} \nonumber\\
&=& 9.21 \times 10^{13}~{\rm s}^{-1} \left(\frac{N_l}{226}\right)\left(\frac{\overline{E_{ul}}}{218~{\rm eV}}\right)
\left(\frac{T_9}{0.4}\right) \nonumber\\
&&\times  \left(\frac{Z_A}{4}\right)^2 \left(\frac{m_A}{6.53~{\rm GeV}}\right)^{-1},
\label{eq23}
\end{eqnarray}
where 
$\overline{E_{ul}}\equiv \sum_l [(g_l/g_u)f_{lu} E_{ul}]/N_l$ was defined.

This stimulated emission rate as well as the spontaneous emission rate is larger for states with smaller main quantum numbers.  The rate for the excitation via the photo-absorption, $\Gamma^{\gamma}_{l,~{\rm ab}}$ is roughly the same as that for the deexcitation.

\subsubsection{Transitions to the GS $^7$Be$_X$}\label{sec264}
The excited states $^7$Be$_X^\ast$ with $n\gtrsim 113$ produced via the charge exchange reaction have binding energies much less than the temperature at the $^7$Be$_X$ destruction epoch of $T_9\gtrsim 0.4$.  The states, therefore, predominantly transit to other states via CBR-absorptions and the stimulated emission by CBR.  Their rates are larger for large $E_{ul}$ values as indicated in Eq. (\ref{eq23}).  The rough assumption of $E_{ul} \sim |E_u| \propto 1/n^2$ leads to a scaling of $\Gamma^{\gamma}_{{\rm Be}_X^\ast,~{\rm st}}\propto 1/n^3$.  This means that deexcitation rates of low energy states with smaller $n$ values are larger than those of high energy states.  Therefore, initially produced states of $^7$Be$_X^\ast$ with $n\gtrsim 113$ can quickly start transitions to lower states, while transitions between higher states are significantly hindered.  The photo-excitation and photo-deexcitation result in an effective population of lower states and an slow population of higher states.

When the excited states with $E_{\rm B}(n)\gtrsim T$ corresponding to $n\lesssim 9(T_9/0.4)^{-1/2}$ are formed via the deexcitation, they would transit gradually to the GS through mainly the spontaneous emission (Sec. \ref{sec261}).  The spontaneous emission rate $\propto 1/n^3$ [Eq. (\ref{eq15})] of the lower states is much larger than that for states with $n\gtrsim 113$.  Time scale for the successive reactions producing the GS is given by $\sim {\Gamma^{\gamma}_{{\rm Be}_X^\ast,~{\rm st}}}^{-1}$ since reaction time scales of lower energy states are shorter, and can be neglected.  In this work, we do not treat more details on such transitions to the GS, and assume that the transition rate is given by 
\begin{equation}
\Gamma_{{\rm Be}_X^\ast,~{\rm tr}}=\Gamma^{\gamma}_{{\rm Be}_X^\ast,~{\rm st}}.
\label{eq29}
\end{equation}

Transitions to the $^7$Be$_X$ GS can realize through multiple stimulated emissions and photo-absorptions induced by CBR and spontaneous emissions.  If the destruction rate of an excited state $u$, $\Gamma_{u,~{\rm de}}$, is larger than its transition rate $\Gamma_{u,~{\rm tr}}$, the probability for transitions through multiple ($N$ times) photo-emissions $\sim (\Gamma_{u,~{\rm tr}}/\Gamma_{u,~{\rm de}})^N$ becomes very small.  In this case, therefore, an efficient formation of the $^7$Be$_X$ GS prefers a direct production of lower energy states at the $^7$Be$_X^\ast$ formation reaction or their production at the first photo-emission after the formation of $^7$Be$_X^\ast$ excited states.  The first case is, however, not expected to operate effectively by an analogy of protonium formation dominantly producing states of large main quantum numbers $n\gtrsim n_{\rm eq}$ which correspond to small binding energies \cite{Cohen1997}.  

We note that the $^7$Be nucleus has been assumed to be a point charge particle in this work.  Binding energies of $A_X$ atomic GS can, however, be significantly smaller than those estimated in the assumption of point charge nuclei since Bohr radii for the $A_X$ are of nuclear dimensions $\sim \mathcal{O}$(fm) \cite{cahn:1981,Bird:2007ge,Kusakabe:2007fv}.  Resultingly, the radiative dipole transition rate of $^7$Be$_X^\ast$ to the GS, which scales with the third power of emitted photon energy, is significantly smaller than that estimated for the point charge nucleus.  However, spontaneous emission rates of low energy levels $n\sim \mathcal{O}(1)$ are much larger than those of high energy levels $n\sim \mathcal{O}(10-100)$.  The total time scale for the transition to the GS is, therefore, contributed mainly from transitions of the highly excited states of $^7$Be$_X^\ast$.  The excited states have Bohr radii much larger than the charge radius of $^7$Be, and their binding energies are almost the same as those estimated for a point charge $^7$Be.  The effect of a finite size nuclear charge is, therefore, not relevant in this transition rate.

\subsection{$^7$Be$_X^\ast$ destruction}\label{sec27}
The main reaction of $^7$Be$_X^\ast$ destruction is the ionization at collisions with electron and positron, i.e., $^7$Be$_X^\ast$+$e^\pm\rightarrow ^7$Be+$X^-$+$e^\pm$.  There is no study about the cross section for this type of reaction.  We then assume the cross section by an analogy of cross sections for ionizations of normal hydrogen which has been measured, and of muonic hydrogen-like particle consisting of a muon and a proton \cite{Kunc1994,Kunc1998}.  The energy threshold for the ionization of $^7$Be$_X^\ast$ is determined from the condition that an energy transferred from $e^\pm$ to $^7$Be at a collision be larger than ionization potential of $^7$Be$_X^\ast$ as $E_{\rm th}=(m_{^7{\rm Be}}/m_e)I(^7{\rm Be}^{3+})/8=348$
 keV $(n/113)^{-2}$~\cite{Kunc1994}.  The cross section for energies above the threshold nearly corresponds to the Bohr radius squared, i.e., $\pi\{2n^2/[Z_{^7{\rm Be}}\alpha \mu(^7{\rm Be}, X^-)]\}^2$, (Fig. 4 of Ref. \cite{Kunc1994}; curve 2).  We roughly assume that the destruction cross section of $^7$Be$_X^\ast$ is $\sigma_{\rm de}(E)=\sigma_{\rm de}=\pi \{2n^2/[Z_{^7{\rm Be}}\alpha \mu(^7{\rm Be}, X^-)]\}^2$ independent of energy.  Using Eq. (\ref{eq10}) we find
\begin{eqnarray}
\langle \sigma v \rangle &=&\left[\frac{8T}{\pi \mu(^7{\rm Be}_X^\ast, e^-)}\right]^{1/2} \left(1+\frac{E_{\rm th}}{T}\right) \exp\left(-\frac{E_{\rm th}}{T}\right) \sigma_{\rm de}\nonumber\\
&\approx&1.66
n^4 \times 10^{-17}~{\rm cm}^3~{\rm s}^{-1} \left(\frac{T_9}{0.4}\right)^{1/2} \left(1+\frac{E_{\rm th}}{T}\right) \nonumber\\
&&\times \exp\left(-\frac{E_{\rm th}}{T}\right) \left(\frac{\sigma_{\rm de}}{1.35
~n^4~{\rm mb}}\right).
\label{eq16}
\end{eqnarray}
We note that the equation $E_{\rm th}/T\ll1$ is satisfied for $n\gg1$.

The rate of the Be$_X^\ast$ destruction through the collisional ionization is then given by
\begin{eqnarray}
\Gamma_{{\rm Be}_X^\ast,~{\rm de}}&=&(n_{e^-}+n_{e^+}) \langle \sigma v \rangle\nonumber\\
&=&2.79
\times 10^{10}~{\rm s}^{-1}~\left(\frac{T_9}{0.4}\right)^2 \frac{\mathrm{e}^{-m_e/T}}{2.97\times 10^{-7}} \nonumber\\
&&\times \frac{(1+E_{\rm th}/T)}{11.1
} \frac{\mathrm{e}^{-E_{\rm th}/T}}{4.13
\times 10^{-5}} \left(\frac{n}{113}\right)^4,
\label{eq17}
\end{eqnarray}
where
Eqs. (\ref{eq3}) and (\ref{eq16}) were used, and $T_9=0.4$ and $n=113$ (corresponding to the GS \beIV) was assumed for numerical values.  The rate depends on the temperature exponentially, and is larger at higher temperature.  The rate also significantly depends on the main quantum number through the variable $E_{\rm th}\propto n^{-2}$ and as the power of $n^4$.  Rates for smaller $n$ values are smaller because of smaller values of $\mathrm{e}^{-E_{\rm th}/T}$.

We can regard excited states of $^7$Be$_X^\ast$ as effective paths to the GS if the transition rates are larger than their destruction rates.  This treatment was utilized in an estimation of relic abundances of strongly interacting massive particles (SIMPs) after the cosmological color confinement by calculations for the annihilation of massive colored particles ($Y$'s) through a formation of resonant or bound states of $(Y\bar{Y})^\ast$ \cite{Kusakabe:2011hk}.  From a comparison of rates for the $^7$Be$_X^\ast$ destruction [Eq. (\ref{eq17})] and the transition to the GS [Eqs. (\ref{eq29}) and (\ref{eq23})], we find that excited states of $^7$Be$_X^\ast$ with $n \gtrsim 113$ produced via the charge exchange reaction of the GS \beIV\ is the effective path in the temperature range of $T_9\gtrsim 0.4$, while those with $n\gtrsim 226$ produced via the reaction of excited state \beIV\ with $n\geq 2$ are not.  As will be shown below, the recombination through the charge exchange is insignificant compared with the radiative recombination for $T_9 \lesssim 0.4$.  Since the GS \beIV\ is the only available path in the epoch of $T_9\gtrsim 0.4$, we take into account only the GS \beIV\ in this work.

\subsection{$^7$Be$_X^\ast$ charge exchange}\label{sec29}
In this subsection, we show that the $^7$Be charge exchange reaction, $^7$Be$_X^\ast +e^- \rightarrow ^7$Be$^{3+} +X^-$, is not an effective destruction process.  Below, the rate for destruction via the charge exchange $\Gamma_{{\rm Be}_X^\ast \rightarrow {\rm Be}^{3+}}$ is estimated and found to be smaller than the spontaneous emission rate, $\Gamma_{u,~{\rm sp}}$, or the destruction rate for $e^\pm$ collisional ionization, $\Gamma_{{\rm Be}_X^\ast,~{\rm de}}$.

The reaction $^7$Be$_X^\ast + e^- \rightarrow$\beIV$ +X^-$ is the inverse of the reaction for $^7$Be$_X^\ast$ production.  In thermal environment, the inverse reaction rate $\langle \sigma v\rangle_{\rm i}$ is related to the forward reaction rate $\langle \sigma v\rangle_{\rm f}$ through the detailed balance \cite{Rybicki1979} as
\begin{eqnarray}
\langle \sigma v\rangle_{\rm i} &=&\frac{g_{^7{\rm Be}^{3+}} \cdot g_X}{g_e \cdot g_{^7{\rm Be}_X^\ast}} \left(\frac{m_{^7{\rm Be}^{3+}} \cdot m_X}{m_e \cdot m_{^7{\rm Be}_X^\ast}}\right)^{3/2} \mathrm{e}^{-Q/T} \langle \sigma v\rangle_{\rm f} \nonumber\\
&\approx& \frac{1}{2\ell+1} \left(\frac{m_{^7{\rm Be}}}{m_e}\right)^{3/2} \langle \sigma v\rangle_{\rm f}
\label{eq30}
\end{eqnarray}
where
$Q$ is the reaction $Q$ value, and
$\ell$ is the azimuthal quantum number of $^7$Be$_X^\ast$.
Since the amplitude of $Q$ value is of the order of the binding energy of \beIV\ at most, the condition $|Q|/T \ll 1$ is satisfied.  

The rate of Eq. (\ref{eq11}) for the reaction \beIV$+X^- \rightarrow ^7$Be$_X^\ast +e^-$ is a total rate as a sum over rates for multiple final states.  Rates for respective final states are, therefore, smaller than the total rate.  As found in studies on the protonium formation \cite{Cohen1997}, many final states of $^7$Be$_X^\ast$ with different main and azimuthal quantum numbers are produced.  We assume that the effective number $N_{\rm f}$ of final states are produced with roughly equal partial cross sections.  Then, the forward rate for one final state $\langle \sigma v \rangle_{\rm f}$ is given by Eq. (\ref{eq11}) divided by $N_{\rm f}$.  The inverse reaction rate is then estimated as
\begin{eqnarray}
\Gamma_{{\rm Be}_X^\ast \rightarrow {\rm Be}^{3+}}&=&n_{e^-} \langle \sigma v\rangle_{\rm i} \nonumber \\
&=& \frac{4}{3\pi^2} m_{^7{\rm Be}} I(^7{\rm Be}^{3+})^2 \frac{\sigma(I(^7{\rm Be}^{3+}))}{N_{\rm f}(2\ell+1)} \mathrm{e}^{-m_e/T} \nonumber \\
&=& 8.68\times 10^9~{\rm s}^{-1} \left(\frac{N_{\rm f}}{10^3}\right)^{-1} \left(\frac{2\ell+1}{100}\right)^{-1} \nonumber\\
&&\times \left[\frac{\sigma(I(^7{\rm Be}^{3+}))}{1.75 \times 10^7~{\rm b}}\right] \left(\frac{\mathrm{e}^{-m_e/T}}{2.97\times 10^{-7}}\right),
\label{eq31}
\end{eqnarray}
where
Eqs. (\ref{eq3}), (\ref{eq11}), and (\ref{eq30}) were used, and it was assumed that states of $^7$Be$_X^\ast$ with 30 different main and azimuthal quantum numbers, respectively, are produced in the forward reaction, i.e., $N_{\rm f}\sim 10^3$.  This charge exchange rate is smaller than the destruction rate via the $e^\pm$ collisional ionization at $T_9\gtrsim 0.4$, where the destruction significantly affect the effective rate of the $^7$Be$_X$ formation through \beIV\ ion [cf. Eq. (\ref{eq18})].  In addition, this rate is smaller than rates of spontaneous and stimulated emissions at $T_9 \lesssim 0.4$ so that it is not an effective destruction process.  The charge exchange reaction is then safely neglected.

\subsection{$^7$Be$_X^\ast$ photoionization}\label{sec28}
In this subsection, we show that the ionization of $^7$Be$_X^\ast$ by CBR is not an effective destruction process.  Below, the rate for destruction via photoionization $\Gamma^{\gamma-{\rm ion}}_{{\rm Be}_X^\ast}$ is estimated and found to be always smaller than the spontaneous emission rate $\Gamma_{u,~{\rm sp}}$ in the relevant temperature region of $T_9 \gtrsim 0.4$.

Cross sections of atomic photoionization have the following characteristics \cite{Karzas1961}.  When the kinetic energy in the final scattering state is smaller than the atomic binding energy, i.e., $E\lesssim E_{\rm B}(n) = Z^2 \alpha^2 \mu/(2n^2)$, the cross section is approximately given by Kramer's semi-classical cross section,
\begin{equation}
\sigma_{nl\rightarrow E}^{\rm K}(E_\gamma)= \frac{2^5 \pi}{3^{3/2}} \frac{Z^2 \alpha}{\mu E_\gamma} \frac{1}{n} \left( \frac{\rho^2} {1+\rho^2}\right)^2,
\label{eq19}
\end{equation}
where
$E_\gamma=E_{\rm B}(n)+E$ is the energy of ionizing photon, 
$\rho=[E_{\rm B}(n)/E]^{1/2}$ \cite{Karzas1961} is defined, and
the index of the charge number $Z$ and arguments of the reduced mass $\mu$ are omitted.  This cross section is different from that for normal hydrogen-like electronic ion by the factor of $Z^2$.  In the current assumption that the mass of $X^-$ is much larger than that of $^7$Be, the effective charge for the electric dipole transition in the center of mass system \cite{Bertulani:2003kr} is $e_1=(m_X Z_{^7{\rm Be}} - m_{^7{\rm Be}} Z_X)e/(m_{^7{\rm Be}}+m_X)\sim Z_{^7{\rm Be}}e=4e$.  On the other hand, the effective charge in the electronic ion is necessarily $e_1\sim -e$.  This difference is reflected in the above equation.

The photoionization rate is given by
\begin{eqnarray}
\Gamma^{\gamma-{\rm ion}}_{{\rm Be}_X^\ast}&=&n_\gamma \langle \sigma_{nl}^{\rm K} \rangle \nonumber \\
&=& \frac{1}{\pi^2} \int_0^\infty \frac{E_\gamma^2}{\exp(E_\gamma/T)-1} \sigma_{nl\rightarrow E}^{\rm K}(E_\gamma)~dE_\gamma \nonumber\\
&=&\frac{2^5}{3^{3/2} \pi} \frac{Z^2 \alpha}{\mu n} \int_0^\infty \frac{E_\gamma^2}{\exp(E_\gamma/T)-1} \left( \frac{\rho^2} {1+\rho^2}\right)^2 dE_\gamma. \nonumber\\
\label{eq20}
\end{eqnarray}

The integral in Eq. (\ref{eq20}) can be approximately estimated as follows.  We divide the integral interval to (1) $E\lesssim E_{\rm B}(n)$ and (2) $E_\gamma \sim E \gtrsim E_{\rm B}(n)$.  As for component (1), we assume $[\rho^2/(1+\rho^2)]^2 \sim 1$, and $\exp(E_\gamma/T)-1 \sim E_\gamma/T$ because of $E_\gamma/T \ll 1$.  As for component (2), we assume $[\rho^2/(1+\rho^2)]^2 \sim \rho^4$, and $\exp(E_\gamma/T)-1 \sim E_\gamma/T$ since the region of $E_\gamma/T \ll 1$ gives the dominant contribution to the integral.  Under these approximation, both components of integral are found to be $\sim E_{\rm B}(n)T$.  We then find the photoionization rate from the order of magnitude estimation,
\begin{eqnarray}
\Gamma^{\gamma-{\rm ion}}_{{\rm Be}_X^\ast} &\sim & \frac{2^6}{3^{3/2} \pi} \frac{Z^2 \alpha}{\mu n} E_{\rm B}(n) T =\frac{2^5}{3^{3/2} \pi} \frac{Z^4 \alpha^3}{n^3} T\nonumber\\
&=&7.07 \times 10^9~{\rm s}^{-1}~\left(\frac{Z}{4}\right)^4 \left(\frac{n}{113}\right)^{-3} \nonumber\\
&&\times \left(\frac{T_9}{0.4}\right).
\label{eq21}
\end{eqnarray}
This photoionization rate as well as the spontaneous emission rate scales as $\propto 1/n^3$, and is larger for states with smaller main quantum numbers.  The photoionization rate of $^7$Be$_X^\ast$ is always much smaller than the spontaneous emission rate for excited states with $n\lesssim 100$ at $T_9\gtrsim 0.4$.  The photoionization, therefore, never becomes an effective destruction process [cf. Eq. (\ref{eq18})], and is neglected.

\section{results}\label{sec3}

We show an effect of the nonradiative recombination on BBN.

\begin{figure}
\begin{center}
\includegraphics[width=8.0cm,clip]{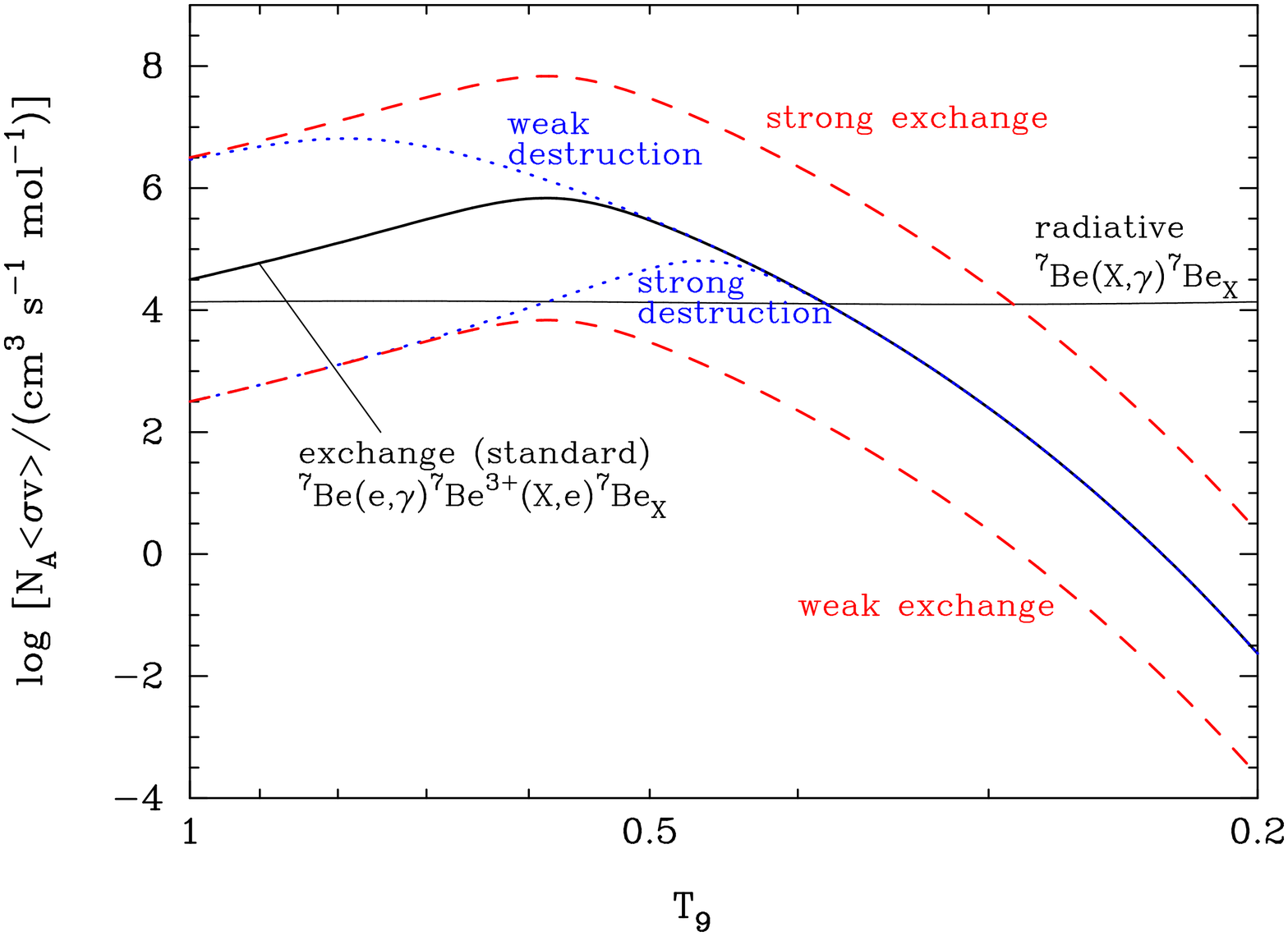}
\caption{(color online).  Effective recombination rates for $^7$Be($e^-$, $\gamma$)$^7$Be$^{3+}$($X^-$, $e^-$)$^7$Be$_X$ (thick lines) and direct radiative capture rate for $^7$Be($X^-$, $\gamma$)$^7$Be$_X$ (thin solid line) as a function of temperature $T_9$.  The thick solid line shows the standard rate estimated with Eqs. (\ref{eq18}), (\ref{eq6}), (\ref{eq13}), (\ref{eq23}), (\ref{eq29}), and (\ref{eq17}).  Upper and lower dashed lines correspond to rates in which the cross section of charge exchange reaction $^7$Be$^{3+}$($X^-$, $e^-$)$^7$Be$_X^\ast$, $\sigma(I(^7{\rm Be}^{3+}))$, is multiplied by 100 and 0.01, respectively.  Lower and upper dotted lines correspond to rates in which the cross section of the $^7$Be$_X^\ast$ destruction reaction, $\sigma_{\rm de}$, is multiplied by 100 and 0.01, respectively. \label{fig1}}
\end{center}
\end{figure}

Figure \ref{fig1} shows effective recombination rates for $^7$Be($e^-$, $\gamma$)$^7$Be$^{3+}$($X^-$, $e^-$)$^7$Be$_X$ (thick lines) and direct radiative capture rate for $^7$Be($X^-$, $\gamma$)$^7$Be$_X$ (Eq. (2.9) of Ref. \cite{Bird:2007ge}; thin solid line) as a function of temperature $T_9$.  For example we assumed that the effective rate is given by Eq. (\ref{eq18}), whose respective terms are described by the rough order of estimates [Eqs. (\ref{eq6}), (\ref{eq13}), (\ref{eq23}), (\ref{eq29}) and (\ref{eq17})].  We fix the physical values except for the $^7$Be$_X^\ast$ formation cross section at $E=I(^7{\rm Be}^{3+})$, $\sigma(I(^7{\rm Be}^{3+}))$, and its destruction cross section, $\sigma_{\rm de}$.  In order to conservatively take into account uncertainties in the magnitudes of cross sections, and also to check dependences of the recombination rate on cross sections,  five sets of cross section values are assumed: ( $\sigma(I(^7{\rm Be}^{3+}))$, $\sigma_{\rm de}$)=(17.5 Mb, 0.219
 Mb) (standard rate: thick solid line), ($1.75\times 10^3$ Mb, 0.219 Mb) and (0.175 Mb, 0.219 Mb) (upper and lower dashed lines), and (17.5 Mb, 21.9 Mb) and (17.5 Mb, $2.19\times 10^{-3}$ Mb) (lower and upper dotted lines).


We used a BBN code including many reactions associated with the $X^-$ particle \cite{Kusakabe:2007fu,Kusakabe:2007fv,Kusakabe:2010cb} constructed with the public Kawano code \cite{Kawano1992,Smith:1992yy}.  We then solve the nonequilibrium nuclear and chemical reaction network associated to the $X^-$ particle with improved reaction rates derived from rigorous quantum many-body dynamical calculations \cite{Hamaguchi:2007mp,Kamimura:2008fx}. The neutron lifetime was updated with $878.5 \pm 0.7_{\rm stat} \pm 0.3_{\rm sys}$~s~\cite{Serebrov:2010sg,Mathews:2004kc} based on improved measurements \cite{Serebrov:2004zf}.  The baryon-to-photon ratio was taken from the value determined by the WMAP \cite{Spergel:2003cb,Spergel:2006hy,Larson:2010gs,Hinshaw:2012fq} for model $\Lambda$CDM (WMAP9 data only):  $\eta=(6.19\pm 0.14) \times 10^{-10}$~\cite{Hinshaw:2012fq}.  The reaction rate for $^4$He$_X$($\alpha$, $\gamma$)$^8$Be$_X$ was updated with that in Ref. \cite{Pospelov:2007js}.  The new reaction suggested in this work, i.e., $^7$Be($e^-$, $\gamma$)$^7$Be$^{3+}$($X^-$, $e^-$)$^7$Be$_X$, was then included.

\begin{figure}
\begin{center}
\includegraphics[width=8.0cm,clip]{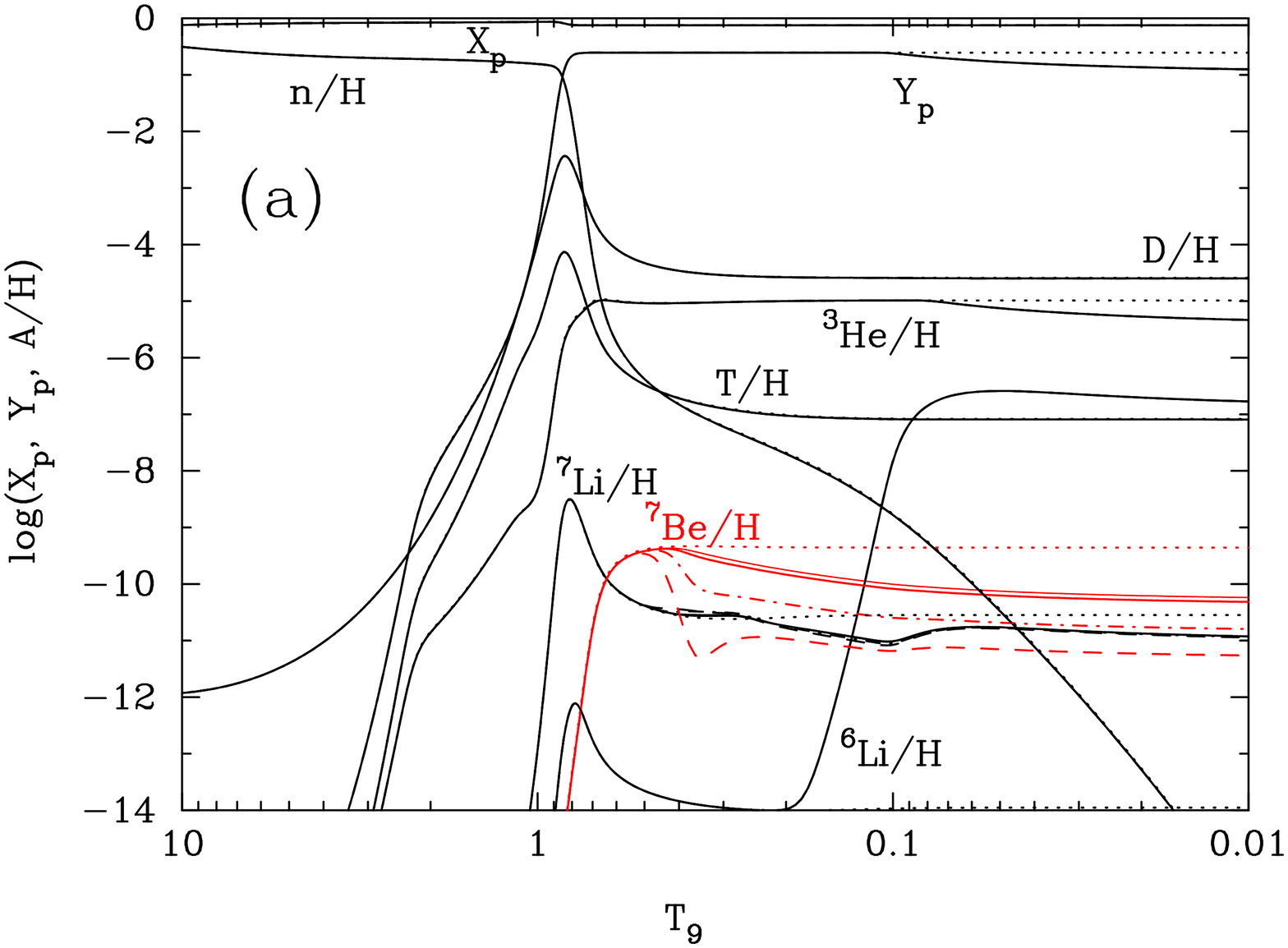}\\
\includegraphics[width=8.0cm,clip]{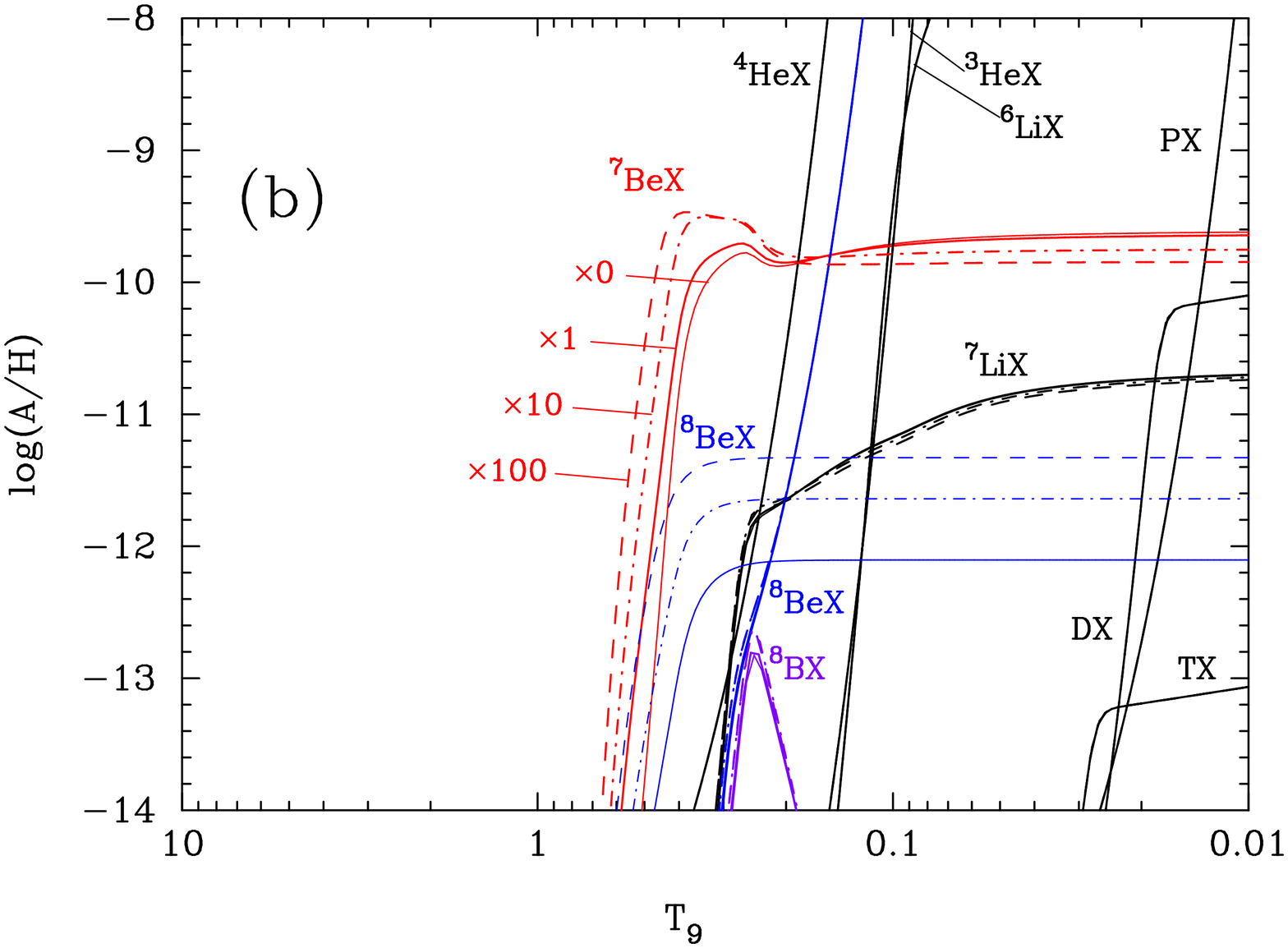}
\caption{(color online).  Calculated abundances of normal nuclei (a) and $X$ nuclei (b) as a function of $T_9$.  The abundance and the lifetime of $X^-$ particle are taken to be $Y_X=n_X/n_b=0.05$ and $\tau_X=\infty$, respectively.  $X_{\rm p}$ and $Y_{\rm p}$ are mass fractions of $^1$H and $^4$He, respectively, while other lines correspond to number abundances with respect to that of $^1$H.  Thick lines correspond to the standard reaction rate for $^7$Be($e^-$, $\gamma$)$^7$Be$^{3+}$($X^-$, $e^-$)$^7$Be$_X$, while thick dot-dashed and dashed lines correspond to the cross sections of $^7$Be$^{3+}$($X^-$, $e^-$)$^7$Be$_X^\ast$ multiplied by a factor of 10 and 100, respectively.  The thin solid lines correspond to the case in which the recombination through $^7$Be$^{3+}$ is completely neglected.  The dotted lines show a result of SBBN calculation.  $^8$Be$_X$ abundances are drawn for both cases with (thick lines) and without (thin lines) the reaction $^4$He$_X$($\alpha$, $\gamma$)$^8$Be$_X$.  \label{fig2}}
\end{center}
\end{figure}

Figure \ref{fig2} shows calculated abundances of normal nuclei (a) and $X$ nuclei (b) as a function of $T_9$.   $X_{\rm p}$ and $Y_{\rm p}$ are mass fractions of $^1$H and $^4$He, respectively, while other lines correspond to number abundances with respect to that of $^1$H.  Nuclear abundances are shown for all nuclei whose abundances are larger than $10^{-14}$ times the H abundance except for $^8$Li.  Thick solid lines correspond to the standard rate for $^7$Be($e^-$, $\gamma$)$^7$Be$^{3+}$($X^-$, $e^-$)$^7$Be$_X$, while thick dot-dashed and dashed lines correspond to the charge exchange cross section $\sigma(I(^7{\rm Be}^{3+}))$ which is higher than that of the standard case by a factor of 10 and 100, respectively.  We have chosen these three cases since the resulting $^7$Be$_X$ abundance is most sensitive to the cross section especially when the cross section is large.  The thin solid line corresponds to the case in which the recombination through $^7$Be$^{3+}$ is completely neglected.  The dotted lines show a result of SBBN calculation.  Depending on the parameter set for cross sections, time evolutions of $^7$Be and $^7$Be$_X$ abundances change.  Therefore, the final abundance of $^7$Be+$^7$Be$_X$ is affected by the efficiency of the recombination of $^7$Be with $X^-$ through the \beIV\ ion.


We note that $^8$Be$_X$ is produced in the temperature region of $T_9\sim 0.4$--$0.3$ through the reaction $^7$Be$_X$($d$, $p$)$^8$Be$_X$ \cite{Kusakabe:2007fv} following the production of $^7$Be$_X$.  There are two operative reactions for $^8$Be$_X$ production: $^7$Be$_X$($d$, $p$)$^8$Be$_X$ and $^4$He$_X$($\alpha$, $\gamma$)$^8$Be$_X$.  In Fig. \ref{fig2}, $^8$Be$_X$ abundances are shown for both cases with (thick lines) and without (thin lines) the reaction $^4$He$_X$($\alpha$, $\gamma$)$^8$Be$_X$.  When the reaction $^4$He$_X$($\alpha$, $\gamma$)$^8$Be$_X$ is switched off, $^8$Be$_X$ is found to be produced via $^7$Be$_X$($d$, $p$)$^8$Be$_X$, and the $^7$Be abundance changes because of the variable recombination rates.  However, in this epoch, the reaction $^8$Be$_X$($\gamma$, $\alpha$)$^4$He$_X$, that is the inverse reaction of the latter reaction above, operates effectively as the destruction reaction of $^8$Be$_X$.  This destruction is caused by a small reaction $Q$ value of the forward $^4$He$_X$($\alpha$, $\gamma$)$^8$Be$_X$ reaction ($Q=0.966$ MeV in the model of Ref. \cite{Kusakabe:2007fv}).  The $^8$Be nuclei is unstable against the $\alpha$ emission so that $^8$Be$_X$ is barely bound with respect to the $\alpha$+$^4$He$_X$ separation channel although it is stabilized by the Coulomb attractive force of $X^-$.  The small $Q$ value means a small energy threshold of the inverse reaction and resultingly a large inverse reaction rate.  At $T_9\sim 0.1$, $^8$Be$_X$ is produced via the reaction $^4$He$_X$($\alpha$, $\gamma$)$^8$Be$_X$ because of an increased $^4$He$_X$ abundance.  Since larger exchange rates for $^7$Be$^{3+}$($X^-$, $e^-$)$^7$Be$_X^\ast$ lead to earlier enhancement of $^7$Be$_X$ abundance, $^8$Be$_X$ production at $T_9\sim 0.4$--$0.3$ proceeds strongly when the rate is larger although its effect does not seen in curves for the case with the reaction $^4$He$_X$($\alpha$, $\gamma$)$^8$Be$_X$.  

\section{summary}\label{sec4}
We studied effects of a long-lived negatively charged massive particle, i.e., $X^-$ on BBN.  In this BBN model including the $X^-$, $^7$Be destruction can occur, and the $^7$Be abundance can reduce from the abundance predicted in SBBN model.  Resultingly the final abundance of $^7$Li can explain the abundances observed in MPSs.  The $^7$Be destruction proceeds through formation of $^7$Be$_X$ followed by its radiative proton capture reaction.  In this work we suggest a new route of $^7$Be$_X$ formation, i.e., the $^7$Be charge exchange between \beIV\ ion and $X^-$.  

What we have found are summarized as follows.

\begin{enumerate}
\item The rate for the $^7$Be$_X$ formation through \beIV\ depends on the number fraction of \beIV\ ion, the charge exchange cross section of \beIV\, and the probability that produced excited states $^7$Be$_X^\ast$ are converted to the GS (Sec. \ref{sec21}).  

\item In the $^7$Be exchange reaction, i.e., $^7$Be$^{3+}$($X^-$, $e^-$)$^7$Be$_X^\ast$, the GS \beIV\ is converted to highly excited states of $^7$Be$_X^\ast$ with main quantum numbers of $n\sim 113$.  The GS \beIV\ and the excited states $^7$Be$_X^\ast$ have almost equal sizes of binding energies and atomic radii (Sec. \ref{sec22}).  

\item In the epoch of $^7$Be destruction in the BBN model including the $X^-$, the \beIV\ ion can be regarded as an isolated system which is separated from particles in thermal bath in the universe (Sec. \ref{sec23}).  

\item In the epoch, the rate for the recombination, $^7$Be$^{4+}+e^-\rightarrow$ \beIV\ $+\gamma$, is large enough, and the equilibrium abundance ratio of the fully ionized $^7$Be$^{4+}$ and the partially ionized \beIV\ had been realized (Sec. \ref{sec24}).  

\item The cross section of the charge exchange reaction $^7$Be$^{3+}$($X^-$, $e^-$)$^7$Be$_X^\ast$ is estimated from an analogy of the protonium ($p\bar{p}$) formation.  The rate for \beIV\ to form $^7$Be$_X^\ast$ through this reaction is then calculated.  The rate is proportional to the $X^-$ abundance $Y_X$, and can be much larger than the cosmic expansion rate in the epoch of the $^7$Be destruction (Sec. \ref{sec25}).  

\item The bound-bound transition rate of $^7$Be$_X^\ast$ produced via this charge exchange reaction to the GS $^7$Be$_X$ is estimated by applying the electric dipole transition rate to the exotic atomic system of $^7$Be$_X^\ast$ (Sec. \ref{sec26}).  

\item The cross section of the $^7$Be$_X^\ast$ destruction via $e^\pm$ collisional ionization is estimated from an analogy of the destruction of muonic hydrogen, $\mu p$, at a collision with $e^-$.  The rate for $^7$Be$_X^\ast$ destruction through this reaction is then calculated.  The rate is smaller than that for the transition to the GS only for excited states $^7$Be$_X^\ast$ with main quantum number $n\gtrsim 113$.  Such excited states are produced via the $^7$Be exchange of the GS \beIV.  The GS \beIV\ is, therefore, the only effective path of the GS $^7$Be$_X$ formation (Sec. \ref{sec27}).  

\item The $^7$Be charge exchange rate of $^7$Be$_X^\ast$ via the $e^-$ collision is estimated with the detailed balance relation.  The rate is smaller than the destruction rate via the $e^\pm$ collisional ionization at $T_9\gtrsim 0.4$, while it is smaller than rates of spontaneous and stimulated emissions at $T_9 \lesssim 0.4$.  The charge exchange reaction, therefore, does not work as an effective reaction for $^7$Be$_X^\ast$ destruction (Sec. \ref{sec29}).

\item The photoionization rate of $^7$Be$_X^\ast$ is estimated with a classical cross section.  Differently from the rate for the destruction in electronic collisional ionization, the photoionization rate is always much smaller than the spontaneous emission rate.  The photoionization is, therefore, not an important reaction of $^7$Be$_X^\ast$ destruction (Sec. \ref{sec28}).  

\item Using physical quantities relevant to the $^7$Be$_X$ formation through \beIV\ estimated in this work, effective recombination rates are derived as a function of cosmic temperature for several example cases.  Our primary rate for the $^7$Be$_X$ formation through \beIV\ is larger than the rate for the direct formation via the radiative recombination of $^7$Be and $X^-$.  Uncertainties in cross sections of the charge exchange $^7$Be$^{3+}$($X^-$, $e^-$)$^7$Be$_X^\ast$ and the destruction $^7$Be$_X^\ast$($e^\pm$, $e^\pm$ $^7$Be)$X^-$ affect the $^7$Be$_X$ formation rate.  The importance of this reaction is shown in BBN calculations with a latest nonequilibrium reaction network code (Sec \ref{sec3}).
\end{enumerate}

This study shows a possibility that the $^7$Li problem is caused predominantly by the new reaction pathway through  $^7$Be$^{3+}$.

\begin{acknowledgments}
We are grateful to Professor Masayasu Kamimura for discussion on the reaction cross sections.  This work was supported by the National Research Foundation of Korea (Grant Nos. 2012R1A1A2041974, 2011-0015467, 2012M7A1A2055605), and in part by Grants-in-Aid for Scientific Research of JSPS (24340060), and Scientific Research on Innovative Area of MEXT (20105004).

\end{acknowledgments}


%

\end{document}